\documentclass[aps,nofootinbib,prl,reprint,superscriptaddress]{revtex4-1}

\pdfoutput=1
\usepackage{graphicx}

\usepackage{multirow}

\usepackage{amsmath,amssymb,mathrsfs}

\usepackage[bookmarks=false]{hyperref} %make sure it comes last of your loaded packages
\hypersetup{pdfstartview=FitH,pdfhighlight=/O,colorlinks=false}

\newcommand{\Reals}{{\mathbb R}}
\newcommand{\xvec}{{\bf x}}
\newcommand{\pvec}{{\bf p}}
\newcommand{\del}{\nabla}
\newcommand{\sgn}{\mathop{\rm sgn}\nolimits}
\newcommand{\tr}{\mathop{\rm tr}\nolimits}
\newcommand{\Jvec}{{\bf J}}
\newcommand{\ket}[1]{\vert #1 \rangle}
\newcommand{\braket}[2]{\langle #1 \vert #2 \rangle}
\newcommand{\Avec}{{\bf A}}

\begin{document}

\title{Maslov indices, Poisson brackets, and singular differential forms}

\author{Ilya Esterlis} 
\affiliation{Perimeter Institute for
Theoretical Physics, 31 Caroline Street North Waterloo, Ontario
Canada N2L 2Y5} 
\email{iesterlis@perimeterinstitute.ca}

\author{Hal M. Haggard}
\affiliation{Aix Marseille Universit\'e, CNRS, CPT, UMR 7332, 13288 Marseille, France.\\ Universit\'e de Toulon, CNRS, CPT, UMR 7332, 83957 La Garde, France.}
\email{haggard@cpt.univ-mrs.fr}

\author{Austin Hedeman}
\affiliation{Department of Physics, University of California,  
Berkeley, California USA}
\email{ajh38@berkeley.edu; robert@wigner.berkeley.edu} 

\author{Robert G. Littlejohn}
\affiliation{Department of Physics, University of California,  
Berkeley, California USA}
\email{robert@wigner.berkeley.edu}

\begin{abstract}
  Maslov indices are integers that appear in semiclassical wave
  functions and quantization conditions.  They are often notoriously
  difficult to compute.  We present methods of computing the Maslov
  index that rely only on typically elementary Poisson brackets and
  simple linear algebra.  We also present a singular differential
  form, whose integral along a curve gives the Maslov index of that
  curve.  The form is closed but not exact, and transforms by an exact
  differential under canonical transformations. We illustrate the method 
  with the $6j$-symbol, which is important in angular momentum theory and in quantum gravity. 
\end{abstract}

%\begin{flushleft}
%Keywords: 
%\end{flushleft}
%\begin{flushleft}
%PACS: 04.60.Pp
%\end{flushleft}

\maketitle

\emph{Introduction.}---Maslov indices are integers representing phase shifts in semiclassical
expressions for wave functions, matrix elements, and position-space
representations of operators in quantum
mechanics \cite{BerryMount:1972,Gutzwiller:1990,BrackBhaduri:1997}.
They are essential to deriving correct quantization conditions and for
obtaining the correct interference patterns in sums over classical
paths, for example, in periodic orbit
expansions \cite{Gutzwiller:1970,Gutzwiller:1980,
  AuerbachKivelsonNicole:1984, RobbinsLittlejohn:1987,
  CreaghLittlejohnRobbins:1990, EckartWintgen:1991,
  SadovskiiShawDelos:1995, HaggertyEtal:1998, Sugita:2000,
  RocciaBrack:2008, FoxmanRobbins:1997}.  Maslov indices are responsible for the
zero-point energy of oscillators and are useful in applications from quantum optics to quantum gravity. 

In this paper we present new techniques for calculating the Maslov
index that involve only simple Poisson brackets and linear algebra. 
For example, in the analysis of spin networks we need just the standard
Poisson brackets of the components of angular momenta among
themselves. Further, a linear dependency between the differentials of 
the angular momenta appears at caustics and allows the calculation to proceed without reference to conjugate angles, a significant simplification.   

Our techniques allow us to put the invariance of the Maslov index under
canonical transformations into a neat form.  The notion that
quantization conditions should be invariant under canonical
transformations goes back to Einstein \cite{Einstein:1917}, and has
been an important part of the mathematical literature on the Maslov
index in recent years \cite{deGosson:1997}.

The mathematical literature on the Maslov index is
extensive \cite{GuilleminSternberg:1977, MaslovFedoriuk:1981,
  RobbinSalamon:1993, CappelEtal:1994, deGosson:1997} but difficult to
use for computational purposes in physical problems.  This is the
case, for example, in the asymptotics of the Wigner
$6j$-symbol \cite{Edmonds:1960}, which plays an important role in the setting for the volume operator in loop
gravity \cite{BianchiHaggard:2011}.  The $6j$-symbol has played a
central role in the road to \cite{PonzanoRegge:1968, TuraevViro:1992, Ooguri:1992} 
and conceptual development of \cite{Carlip:1998, Barrett:1998} loop gravity, but
several authors who have studied its asymptotics have either resorted
to numerical methods to compute the Maslov index \cite{Roberts:1999,
  FreidelLouapre:2003} or have given up entirely.  The only successful
calculations of the Maslov index for the $6j$-symbol have
been those that reduced the problem to a one-dimensional
system \cite{SchultenGordon:1975, TaylorWoodward:2005, Gurau:2008}, an
option not available in problems that are intrinsically
multidimensional, such as the
$9j$-symbol \cite{HaggardLittlejohn:2010}.

We also present several relations satisfied by the Maslov
index, including an expression for it in terms of a singular
differential form that is closed but not exact.  We refer to this
differential form as singular because it is expressed in terms of
Dirac delta functions times the differentials of smooth functions. 
This approach unifies the phase contributions of the action and 
the Maslov phase into a single differential form.

We work in the phase space $\Reals^{2n}$ with coordinates
$(\xvec,\pvec)$.  A wave function $\psi(\xvec)$ has a semiclassical
representation as a sum over branches; each branch has a phase
$S(\xvec)/\hbar$.  With the right understandings, this notation covers
energy and other eigenfunctions, time-dependent wave functions,
kernels of operators such as $\langle\xvec\vert A\vert\xvec'\rangle$,
matrix elements in angular momentum theory, periodic orbit
contributions in the Gutzwiller trace formula, and other cases. 

\emph{Derivation and Results.}---The $n$-dimensional manifold $\pvec=\del S(\xvec)$ in phase space is a
Lagrangian manifold \cite{Arnold:1989}, call it $L$.  It is the level
set $H_i(\xvec,\pvec)=h_i$, $i=1,\ldots,n$, where the $H_i$ are a set
of functions and the $h_i$ their values, and where $\{H_i,H_j\}=0$ on
$L$.  (For energy eigenfunctions, one of the $H_i$ is the
Hamiltonian.)  These Poisson brackets are required to vanish only on
$L$, not necessarily elsewhere in phase space; this means that
conjugate (e.g., angle) variables on $L$ may not exist, a
case that must be covered for applications to angular momentum theory.

Consider a point $(\xvec,\pvec)$ on $L$, and let the Hamiltonian
vector fields generated by the $H_i$ at this point be 
	\begin{equation}
         X_i = \sum_j
        E_{ji}\frac{\partial}{\partial x_j} 
        + F_{ji}\frac{\partial}{\partial p_j},
	\label{vectorsXi}
	\end{equation}
where $E_{ij} = \{x_i,H_j\} = \partial H_j/\partial p_i$, $F_{ij} =
\{p_i,H_j\} = -\partial H_j/\partial x_i$.  Vectors $X_i$ are tangent
to $L$ and span its tangent space, since $\{H_i,H_j\}
=X_j(H_i) =0$ (the final expression is the vector field $X_j$ acting
on the scalar $H_i$).  Matrix $E_{ij}$ is the Jacobian of the
projection $\pi_x$ from $L$ to $\xvec$-space, in the bases $X_i$ and
$\partial/\partial x_i$; and $F_{ij}$ is that of the projection
$\pi_p$ from $L$ to $\pvec$-space, in the bases $X_i$ and
$\partial/\partial p_i$.  Caustics in $\xvec$-space, that is, of the
wave function $\psi(\xvec)$, occur where $E_{ij}$ is singular; here
the semiclassical wave function suffers a phase shift given by
$e^{-im\pi/2}$, where the integer $m$ is the Maslov index.  We view
$m$ as a function of a directed path $\gamma$ on $L$, which passes
through an $\xvec$-space caustic.  Similarly, the matrix $F_{ij}$ is
singular at caustics in $\pvec$-space (that is, caustics of the
momentum space wave function, the Fourier transform of $\psi(\xvec)$).
In one dimension, $p$-space caustics never occur at an $x$-space
caustic.  In higher dimensions, a $\pvec$-space caustic can occur on
top of an $\xvec$-space caustic, that is, $F$ can be singular when $E$
is singular, a case that must be covered in practice.  Initially,
however, we assume that $F$ is nonsingular in a neighborhood of an
$\xvec$-space caustic, in which our curve $\gamma$ lies.

Maslov's method \cite{MaslovFedoriuk:1981} for computing his index
involves switching to the momentum representation in a neighborhood of
the $\xvec$-space caustic.  The phase of the momentum-space wave
function is ${\tilde S}(\pvec)/\hbar$, where ${\tilde S}(\pvec) =
S(\xvec) - \xvec\cdot\pvec$.  Here $\xvec$ is understood to be a
function of $\pvec$ by restricting $(\xvec,\pvec)$ to be on $L$.  The
momentum-space action satisfies $\partial {\tilde S}/\partial p_i =
-x_i$ and $T_{ij} = \partial^2 {\tilde S}/\partial p_i \partial p_j =
-(\partial x_i/\partial p_j)_H = T_{ji}$, where the subscript $H$
indicates that the $H_i$ are held constant, that is, the derivative is
taken on $L$.  The momentum-space wave function is nonsingular (it has
no caustics) in the neighborhood of the $\xvec$-space caustic, but
when we Fourier transform back to the $\xvec$-representation, there is
a phase difference when the integral is evaluated on the two sides of
the $\xvec$-space caustic.  This gives rise to a relative phase shift
in the $\xvec$-space wave function of $e^{-im\pi/2}$, where $m$ is
related to the change in the signature of matrix $T$ by
$m=-(1/2)\Delta\sgn T$.  (The signature is the number of positive
minus the number of negative eigenvalues; $T$ is symmetric and has
real eigenvalues.)  

One or more of the eigenvalues of $T$ pass through
0 at the caustic, that is, $T$ is singular at the caustic.  This can
be seen by expressing $T$ in terms of matrices $E$ and $F$; the
relation is $T=-EF^{-1}$, as can be proved by manipulating partial
derivatives.  Since $F$ is nonsingular in the neighborhood of the
$\xvec$-space caustic (by our assumptions) and $E$ is singular at the
caustic, $T$ is singular at the caustic.  Thus we have $m=(1/2)\Delta\sgn
(EF^{-1}) = (1/2)\Delta\sgn (F^TE)$, where in the final expression we
have used Sylvester's theorem on the invariance of the signature under
congruency transformation by a nonsingular matrix (in this case, $F$)
and where $F^T$ is the transpose of $F$.  Note that $F^TE$, like $T$,
is symmetric.

For simplicity we assume that only one eigenvalue of $T$ (hence of
$F^TE$) changes sign at the caustic; this is the generic situation.
Let $\lambda$ be this eigenvalue of $F^TE$, and let $v$ be the
corresponding (nonzero) eigenvector, so that $F^TEv=\lambda v$.  Also
let $u=Fv$; since $F$ is nonsingular, $u\ne0$.  We consider $F$, $E$,
$\lambda$, $u$ and $v$ to be functions of a parameter $t$ (not
necessarily time) along the curve $\gamma$, and we let $t=0$ at the
caustic, so that $\lambda(0)=0$.  Then $m=\sgn {\dot\lambda}(0)$; we
assume ${\dot\lambda}(0)\ne0$ (the generic situation).  

At $t=0$, $F^TEv=0$; but since $F$ is nonsingular, this implies
$Ev=0$, and $v$ spans the kernel of $E$ at the caustic.  Matrix $E$ is
not symmetric, so its left and right eigenvectors are not transposes
of each other, but since $F^TE$ is symmetric, we have $v^TF^TE=u^TE=0$
at $t=0$, so $u$ spans the (left) kernel of $E$ at $t=0$.  Now by
differentiating $u^TEv = \lambda v^Tv$ with respect to $t$ and using 
$Ev=0$, $u^TE=0$ and $\lambda=0$ at $t=0$, we find
	\begin{equation}
	m=\sgn u^T{\dot E}v,
	\label{theresult}
	\end{equation}
evaluated at $t=0$.  This is our main result for a local calculation
of the Maslov index, that is, in a neighborhood of a caustic.

To calculate $m$ we first find the caustics, which are the places
where $Ev=0$ has a solution $v\ne0$; these are the places where $\det
E=0$.  The matrix $E$ is needed for the amplitude of the semiclassical
wave function, which can be expressed as $|\det
E|^{-1/2}$ \cite{Littlejohn:1990,Aquilantietal:2007}; the amplitude
diverges at the caustics.  At the caustic we find vector $v$ with
an arbitrary normalization and phase; as mentioned, we are assuming
that the kernel of $E$ is one-dimensional.

Next we must find the vector $u=Fv$, which spans the left kernel of
$E$.  If we have the matrix $F$ we can just do the matrix
multiplication, but in many applications the Poisson brackets in $F$
involve angle variables and are not easy to compute.  Moreover, in
some cases the Lagrangian manifold is not a member of a foliation and
angle variables do not exist. 

 A different approach that avoids these
difficulties is based on the geometrical meaning of the vectors $v$
and $u$, which emerges if we multiply (\ref{vectorsXi}) by $v_i$ and sum
on $i$.  At the caustic, where $\sum_i E_{ji}\,v_i=0$, this gives
$\sum_i v_i X_i = \sum_j u_j \partial/\partial p_j$, where we
have used $u=Fv$.  Recalling that the $X_i$ are the Hamiltonian vector
fields generated by the $H_i$, we let $Y_i$ be the Hamiltonian vector
fields generated by the $x_i$, that is, we let $Y_i = -\partial/\partial
p_i$.  Then we have $\sum_i v_i X_i = -\sum_i u_i Y_i$.  We see that
there is a linear combination of the $X_i$, that is, a vector tangent
to $L$, that is equal to a linear combination of the $Y_i$, that is, a
vector tangent to the vertical Lagrangian manifold $\xvec={\rm
const}$.  The two Lagrangian planes tangent to the two manifolds at
the caustic have a nontrivial (one-dimensional) intersection.

If we regard the symplectic form $\omega$ at a point of phase space as
a linear map between vectors and covectors, then Hamilton's equations
for the $H_i$ can be written $X_i = \omega^{-1} dH_i$, and, similarly,
$Y_i = \omega^{-1} dx_i$.  Now multiplying the previous relation by
$\omega$, we obtain
        \begin{equation}
	\sum_i v_i \,dH_i = -\sum_i u_i \,dx_i.
	\label{uveqn}
	\end{equation}
This equation allows the vector $u$ to be determined, given the vector
$v$ and the differentials $dH_i$ and $dx_i$ at the caustic.  The
calculation is just linear algebra in the cotangent space at the
caustic.  Finally, let the curve $\gamma$ be an orbit of one of the
$H$'s, say, $H_n$.  Then the $t$-derivative in ${\dot E}$ is a Poisson
bracket, and the Maslov index is $m=\sgn u^T
\{E, H_n\} v$.  The calculation of the Maslov index is reduced
to the calculation of Poisson brackets and linear algebra.

As an example consider the one-dimensional Hamiltonian
$H=p^2/2M+V(x)$, where $H=H_1$ in the notation above.  In a
1-dimensional case such as this we will write $e$ and $f$ for matrices
$E$ and $F$, which now are scalars. Here $e=\{x,H\}=p/M$,
$f=\{p,H\}=-V'(x)$.  The caustics are where $e=0$, that is, $p=0$.
Choosing $v=1$, we have $u=fv = -V'(x)$.  The same result is obtained
from (\ref{uveqn}), that is, $v\,dH = -u\,dx$, since $dH = V'(x)\,dx$ at
the caustic where $p\,dp/M=0$.  Finally, using ${\dot e} =
\{e,H\}=-V'(x)/M$, we have $m=\sgn [V'(x)^2/M]=+1$.  The Maslov index
always increases by 1 at a turning point in a kinetic-plus-potential
problem.

The $6j$-symbol is a less trivial example.  The quantum
mechanics \cite{Edmonds:1960} involves four angular momenta, $\Jvec_r$,
$r=1,\ldots,4$ that act on a product of four carrier spaces with
quantum numbers $j_r$.  Intermediate angular momenta $\Jvec_{12}
=\Jvec_1+\Jvec_2$ and $\Jvec_{23}=\Jvec_2+\Jvec_3$ with quantum
numbers $j_{12}$ and $j_{23}$ are defined.  The $6j$-symbol concerns
the subspace $\sum_{r=1}^4 \Jvec_r = \Jvec_{\rm tot}=0$, upon which
$J_{12}^2$ and $J_{23}^2$ have eigenbases $\ket{j_{12}}$ and
$\ket{j_{23}}$.  The $6j$-symbol is proportional to the orthogonal
matrix connecting these bases,
	\begin{equation}
	\braket{j_{12}}{j_{23}} = {\rm const} \times
	\left\{ \begin{array}{ccc}
	j_1 & j_2 & j_{12} \\
	j_3 & j_4 & j_{23}
	\end{array}\right\}.
	\label{6jdef}
	\end{equation}
The $6j$-symbol involves a quantum dynamical system in which a state
is a vector in the subspace $\Jvec_{\rm tot}=0$.  
%Interesting observables acting on this space are $J_{12}^2$, $J_{23}^2$ and the
%volume $V=\Jvec_1\cdot(\Jvec_2\times \Jvec_3)$.  

A state of the corresponding classical system is a quadrilateral, not
necessarily planar, whose edges are four classical angular momentum
vectors $\Jvec_r$ of fixed lengths $J_r=|\Jvec_r|$, modulo overall
rotations.  The vectors satisfy $\sum_r \Jvec_r=0$.  If vectors
$\Jvec_{12}=\Jvec_1+\Jvec_2$ and $\Jvec_{23}=\Jvec_2+\Jvec_3$ are
drawn in, the quadrilateral becomes Wigner's
tetrahedron \cite{Wigner:1959} with edge lengths $J_r$, $r=1,\ldots,4$
and $J_{12}=|\Jvec_{12}|$ and $J_{23}=|\Jvec_{23}|$.  The
quadrilateral is flexible; changing its shape while holding $J_r$,
$r=1,\ldots,4$ fixed changes the classical state, as well as the
lengths $J_{12}$ and $J_{23}$.  The space of shapes of the
quadrilateral or tetrahedron is a sphere, the phase space of the
system \cite{Kapovich:1996, Aquilantietal:2012,LittlejohnYu:2009}; the Poisson bracket of
any two functions $f$ and $g$ of $\Jvec_r$, $r=1,\ldots,4$ is $\{f,g\}
= \sum_{r=1}^4 \Jvec_r\cdot(\del_r f \times \del_r g)$, where $\del_r
= \partial/\partial\Jvec_r$; this is the standard Poisson bracket for
classical angular momenta.  

Interesting classical observables on this phase space are $J_{12}$,
$J_{23}$ and $V =\Jvec_1\cdot (\Jvec_2\times\Jvec_3)$ (this is six
times the volume of the tetrahedron).  The Hamiltonian flow generated
by $J_{12}$ is a rotation of vectors $\Jvec_1$ and $\Jvec_2$ about the
axis defined by $\Jvec_{12}$, while holding $\Jvec_3$ and $\Jvec_4$
fixed; we call the conjugate angle $\phi_{12}$.  Similary, $J_{23}$
generates rotations of $\Jvec_2$ and $\Jvec_3$ with angle $\phi_{23}$
about axis $\Jvec_{23}$.  These rotations are rigid, relative motions
of two faces of the tetrahedron about their common edge ($J_{12}$ or
$J_{23}$).  If we denote the interior dihedral angles of the
tetrahedron about edges $J_{12}$ and $J_{23}$ by $\alpha_{12}$ and
$\alpha_{13}$, with $0\le\alpha_{12} ,\alpha_{23}\le\pi$, then when
$V>0$ we have $\phi_{12}=\alpha_{12}$ and $\phi_{23}=-\alpha_{23}$;
this is clear from a picture of the tetrahedron.  With a change of
signs for the case $V<0$ the angles $\phi_{12}$ and $\phi_{23}$ lie in
the range $-\pi\le \phi_{12},\phi_{23} <\pi$ on the space of all
tetrahedra.  For semiclassical purposes we set $J_r =
j_r+1/2$ \cite{PonzanoRegge:1968, Aquilantietal:2007}.

In calculating Poisson brackets the vectors $\Avec_{rs}
=\Jvec_r\times\Jvec_s$ are convenient; the magnitude $A_{rs}
=|\Avec_{rs}|$ is twice the area of the face spanned by $\Jvec_r$,
$\Jvec_s$.  We find $\{J_{12},J_{23}\} =-V/J_{12}J_{23}
=dJ_{12}/d\phi_{23} =-dJ_{23}/d\phi_{12}$; $\{V,J_{12}\} =
dV/d\phi_{12} = A_{34} A_{12}\cos\phi_{12}/J_{12}$; and $\{V,J_{23}\}
=dV/d\phi_{23} =-A_{23} A_{14}\cos\phi_{23}/J_{23}$.

To compute the Maslov index of the $6j$-symbol we compare
$\braket{j_{12}}{j_{23}}$ with the energy eigenfunction $\psi(x)
=\braket{x}{H}$, which shows that we should identify $H$ (or $H_1$)
above with $J_{23}$ and $x$ with $J_{12}$.  As for $p$, we identify it
with $-\phi_{12}$ so that $\{x,p\}=1$ goes into $\{J_{12},
-\phi_{12}\}=1$.  The idea is that the Lagrangian manifold is
specified by $J_{23}=j_{23}+1/2={\rm const}$, while $J_{12}$ provides
the representation of the wave function.  The caustics occur when
$e=\{J_{12},J_{23}\}=0$, that is, when $V=0$; these are the flat
tetrahedra.  To obtain $u$ and $v$ we need a relation between
$dJ_{12}$ and $dJ_{23}$.  This may be obtained by differentiating the
Cayley-Menger \cite{PonzanoRegge:1968} or Gram \cite{LittlejohnYu:2009}
matrix, but an approach based on Poisson brackets may be given.  Let
$V$ be considered a function of $J_{12}$ and $J_{23}$.  Then
$\{V,J_{12}\} = \{J_{23},J_{12}\}\,\partial V/\partial J_{23}$, which
combined with the above gives $\partial V/\partial J_{23} =
A_{12}A_{34}J_{23}\cos\phi_{12}/V$.  Similarly, consideration of
$\{V,J_{23}\}$ gives $\partial V/\partial J_{12}$.  The results are
summarized by
	\begin{equation}
	V\,dV = A_{14}A_{23}\cos\phi_{23}\, J_{12}\,dJ_{12} +
                A_{12}A_{34}\cos\phi_{12}\, J_{23}\,dJ_{23}.
	\end{equation}
Now setting $V=0$ to evaluate at the caustic and writing $v\,dJ_{23} =
-u\,dJ_{12}$, we find $u= A_{14}A_{23}J_{12}\cos\phi_{23}$ and
$v=A_{12}A_{34}J_{23}\cos\phi_{12}$.  Finally, defining $\dot e$ by
$\{e,J_{23}\}$ (that is, evaluating the Maslov index along an orbit of
$J_{23}$), we find $\dot e = A_{14} A_{23}\cos\phi_{23} /J_{12}
J_{23}^2$, when evaluated at the caustic.  Then the Maslov
index is $m=\sgn u{\dot e}v= \sgn\cos\phi_{12}$; it is 1 when
$\phi_{12}=0$, and $-1$ when $\phi_{12}=\pi$ (the only two
possibilities for a flat tetrahedron).  Notice that in this
calculation we did not need any Poisson brackets involving the angles
$\phi_{12}$ or $\phi_{23}$.

The result (\ref{theresult}) was derived under the assumption that
$\det F\ne0$ in a neighborhood of the point where $\det E=0$; but it
turns out to be correct even when $\xvec$- and $\pvec$-space caustics
coincide.  Such a coincidence typically occurs on a
Lagrangian manifold of dimension $\ge2$, and in cases of symmetry,
such as central force motion, it may occur everywhere.  When $\det
F=0$ the vector $v$ must be interpreted as any nonzero
vector in the kernel of $E$ at the caustic, not as the eigenvector of
$F^TE$ with eigenvalue 0.  Vector $u$ is still defined as $Fv$, and
can be calculated exactly as above (without the explicit knowledge of
$F$); although $F$ is singular, it turns out that $Fv\ne0$.  Relevant
theorems covering the case when $\det E=0$ and $\det F=0$ are the
following.  First, $\ker E \cap \ker F = \{0\}$; next, $\ker F^T E =
\ker E \oplus \ker F$; and third, $F$ maps $\ker E$ invertibly into
$\ker E^T$.  Thus, the singular $F$ becomes nonsingular when
restricted to $\ker E$.

So far we have presented local results, useful for calculating $m$ in
the neighborhood of a caustic.  Now we present a global result, valid
over the whole Lagrangian manifold.  If we have a function $f$ on a
manifold, then the singular differential form $\delta(f)\,df =
(1/2)d\sgn f$ is the ``counting form'' for the crossings of the surface
$f=0$, that is, $\int_\gamma \delta(f)\,df$ counts the number of times
$\gamma$ crosses the surface going from negative $f$ to positive,
minus the number of crossings the other way.  Since the caustic set on the
Lagrangian manifold occurs where $\det E=0$, we might suspect that
there is a singular differential form $\mu$, such that the integral of
$\mu$ along $\gamma$ gives the Maslov index associated with the curve,
and that $\mu$ involves $\delta(\det E)\, d(\det E)$.  Indeed,
this is the case; we find
	\begin{equation}
	\mu=\sgn\tr(C^T F) \,\delta(\det E)\, \tr(C^T\,dE),
	\label{mudef}
	\end{equation}
where $C$ is the cofactor matrix of $E$.  This result applies only to
first order caustics, where $\dim\ker E=1$; but higher order caustics
can be perturbed into a set of first-order caustics, so they represent
limiting cases of this form.  Note that $\tr(C^T\,dE) = d(\det E)$, so
the counting form for the surface $\det E=0$ is modulated by the
factor $\sgn\tr(C^T F)$.  It can be shown that $\tr(C^T F)$ is never
zero when $\det E=0$, even if $\det F$ is also 0.  Noting that $C$ is
proportional to $u\otimes v^T$, where $v\ne0$ is a vector in $\ker E$
and $u=Fv$, it is easy to derive (\ref{theresult}) from (\ref{mudef}).
We found it easiest to derive (\ref{mudef}) itself as the limit of the
differential of the phase of the complex amplitude determinant in a
coherent state representation, in the limit in which the coherent
state representation becomes the $\xvec$-representation.  The form
$\mu$ is closed but not exact, so its integral along $\gamma$ is
invariant under continuous deformations of path.  

The phase $S(\xvec)$ is the integral of the differential form
$\theta=\pvec\cdot d\xvec$ on $L$; this form is closed but not exact
(in general) on $L$.  By combining this form with the Maslov form
$\mu$, the Bohr-sommerfeld quantization condition can be expressed as 
$\oint (\theta-\frac{\pi}{2} \mu) = 2n\pi$.  In this way the usual action and
the Maslov phase are unified in a single form.    

The quantization condition cannot depend on the representation, that
is, the system of canonical coordinates in which the calculation is
carried out, an idea that goes back to Einstein \cite{Einstein:1917}.
Taking first the one-dimensional case, we let coordinates $(x',p')$ be
related to $(x,p)$ by a linear transformation,
	\begin{equation}
	\left(\begin{array}{c}
	x' \cr p'
	\end{array}\right)=
	\left(\begin{array}{cc}
	A & B \cr
	C & D
	\end{array}\right)
	\left(\begin{array}{c}
	x \cr p
	\end{array}\right),
	\label{linearct}
	\end{equation}
where $A$, $B$, $C$, $D$ are constants and $AD-BC=1$.  Then $e$ and
$f$ transform into $e'$ and $f'$ by the same matrix as $x$ and $p$.
In one dimension (\ref{mudef}) becomes $\mu=(\sgn f)\delta(e)\,de$.
Writing $\mu'=(\sgn f')\delta(e')\,de'$, the Maslov differential forms
in two systems of canonical coordinates are $\mu$ and $\mu'$.  Then we
find that $\mu-\mu' = dK(e,e')$, where $K=(1/2)\sgn(e'Be)$ when
$B\ne0$, and $K=0$ when $B=0$.  That is, $\mu$ transforms by the
addition of an exact differential when the system of canonical
coordinates is changed, so that $\oint \mu$ is invariant.  In these
calculations we use $(d/dx)\sgn(x) = 2\delta(x)$.

We will just cite the analogous results in the multidimensional case.
The Maslov forms in the two systems of canonical coordinates are a
primed and unprimed version of (\ref{mudef}).  In the multidimensional
case a linear canonical transformation is still specified by
(\ref{linearct}), where now $A$, $B$, $C$ and $D$ are $n\times n$
matrices such that the whole $2n\times 2n$ matrix is symplectic (see
Appendix~A of \cite{Littlejohn:1986}).  Under the linear canonical
transformation (\ref{linearct}) $\mu$ transforms by an exact
differential, $\mu-\mu'=dK$, where now $K=(1/2)\sgn(E^T B^{-1} E')$
when $B$ is nonsingular.  Thus $\oint \mu$ around a closed loop is
independent of the canonical coordinates.  Function $K$ is a
kind of $F_1$-type generating function \cite{Goldstein:1980}.
Knowledge of $K$ allows one to easily switch the Maslov phase from one
representation to another.  With slight changes, it can be used to
switch to the coherent state representation, which is popular in
recent applications \cite{PletyukhovEtal:2002, GargStone:2004} and in
approaches based on geometric quantization.

The results presented in this article are of great assistance in
computing the Maslov index in various applications, including the
$9j$-symbol  \cite{HaggardLittlejohn:2010}, which is intrinsically
2-dimensional.  The strength of this method is that it reduces what is usually a
 delicate and lengthy tracking of signs to just two ingredients: 
the calculation of Poisson brackets of the observables that directly define
the wave function and the corresponding Lagrangian manifolds 
(these are also necessary for computing the amplitude \cite{Littlejohn:1990, Aquilantietal:2007}); and a linear algebra calculation that alleviates any need for the introduction of angle coordinates.  We will report on details, extensions, and
applications of the results presented here in future publications. 

HMH thanks Carlo Rovelli and acknowledges NSF support under Grant No. OISE-1159218.

\providecommand{\href}[2]{#2}\begingroup\raggedright\endgroup


\begin{thebibliography}{10}

\bibitem{BerryMount:1972}
M.~V.~Berry and K.~E.~Mount, 
{\em Rep. Prog. Phys.} {\bf 35} (1972) 315.

\bibitem{Gutzwiller:1990}
M.~C.~Gutzwiller, {\em Chaos in Classical and Quantum Mechanics},
Springer-Verlag (New York, 1990).

\bibitem{BrackBhaduri:1997}
%Matthias Brack and Rajat Bhaduri, {\em Semiclassical physics},
M. Brack and R. Bhaduri, {\em Semiclassical physics},
Addison-Wesley (Reading, Massachusetts, 1997).

\bibitem{Gutzwiller:1970} 
M. C. Gutzwiller, {\em J. Math. Phys.} {\bf 11} (1970) 1791; {\bf
12} (1971) 343.

\bibitem{Gutzwiller:1980}
M. C. Gutzwiller, {\em Phys. Rev. Lett.} {\bf 45} (1980) 150.

\bibitem{AuerbachKivelsonNicole:1984}
%Assa Auerbach, S. Kivelson and Denis Nicole, {\em Phys. Rev. Lett.} 
A. Auerbach, S. Kivelson and D. Nicole, {\em Phys. Rev. Lett.} 
{\bf 53} (1984) 411.

\bibitem{RobbinsLittlejohn:1987}
%Jonathan M. Robbins and Robert G. Littlejohn, {\em Phys. Rev. Lett.} 
J. M. Robbins and R. G. Littlejohn, {\em Phys. Rev. Lett.} 
{\bf 58} (1987) 1388.

\bibitem{CreaghLittlejohnRobbins:1990}
%Stephen Creagh, Robert G. Littlejohn and Jonathan Robbins, {\em Phys. 
S. Creagh, R. G. Littlejohn and J. Robbins, {\em Phys. 
Rev. A} {\bf 42} (1990) 1902.

\bibitem{EckartWintgen:1991} 
%Bruno Eckart and Dieter Wintgen, {\em J. Phys. A} {\bf 24} (1991) 4335.
B. Eckart and D. Wintgen, {\em J. Phys. A} {\bf 24} (1991) 4335.

\bibitem{SadovskiiShawDelos:1995}
D. A. Sadovskii, J. A. Shaw and J. B. Delos, {\em Phys. Rev. Lett.} 
{\bf 75} (1995) 2120.

\bibitem{HaggertyEtal:1998}
%M. R. Haggerty, Neal Spellmeyer, Daniel Kleppner, and J. B. Delos, 
M. R. Haggerty, N. Spellmeyer, D. Kleppner, and J. B. Delos, 
{\em Phys. Rev. Lett.} {\bf 81} (1998) 1592.

\bibitem{Sugita:2000}
%Ayumu Sugita, {\em Phys. Lett. A} {\bf 266} (2000) 321.
A. Sugita, {\em Phys. Lett. A} {\bf 266} (2000) 321.

\bibitem{RocciaBrack:2008} 
%J\'er\^ome Roccia and Matthias Brack, {\em Phys. Rev. Lett.} {\bf 100} 
J. Roccia and M. Brack, {\em Phys. Rev. Lett.} {\bf 100} 
(2008) 200408.

\bibitem{FoxmanRobbins:1997}
J.~A.~Foxman and J.~M.~Robbins, {\em J. Phys. A} {\bf 30} (1997) 8187;
{\em Nonlinearity} {\bf 18} (2005) 2775.

\bibitem{Einstein:1917}
A. Einstein, {\em Verh. Dtsch. Phys. Ges.} {\bf 19} (1917) 82.

\bibitem{deGosson:1997}
%Maurice de Gosson, {\em Maslov classes, metaplectic representation, and
M. de Gosson, {\em Maslov classes, metaplectic representation, and
Lagrangian quantization}, Akademie Verlag (Berlin, 1997).

\bibitem{GuilleminSternberg:1977}
V. Guillemin and S. Sternberg, {\em Geometric Asymptotics}, 
Amer. Math. Soc. (Providence, R.~I., 1977).

\bibitem{MaslovFedoriuk:1981}
V. P. Maslov and M. V. Fedoriuk, {\em Semi-Classical Approximation
in Quantum Mechanics}, Reidel (Boston, 1981).

\bibitem{RobbinSalamon:1993}
%Joel Robbin and Dietmar Salamon, {\em Topology} {\bf 32} (1993) 827.
J. Robbin and D. Salamon, {\em Topology} {\bf 32} (1993) 827.

\bibitem{CappelEtal:1994}
%Sylvain E. Cappell, Ronnie Lee and Edward Y. Miller, {\em Comm. Pure. 
S. E. Cappell, R. Lee and E. Y. Miller, {\em Comm. Pure. 
Appl. Math.} {\bf XLVII} (1994) 121.

\bibitem{Edmonds:1960}
A.~R.~Edmonds, {\em Angular Momentum in Quantum Mechanics},
Princeton Univ. Press (Princeton, 1960).

\bibitem{BianchiHaggard:2011}
E.~Bianchi and H.~M.~Haggard
{\em Phys. Rev. Lett.} {\bf 107} (2011) 011301; {\em Phys. Rev. D} 
{\bf 86} (2012) 124010.

\bibitem{PonzanoRegge:1968}
G.~Ponzano and T.~Regge,
{\em Spectroscopy and Group Theoretical Methods in Physics} ed
F.~Bloch et al, North-Holland (Amsterdam, 1968).

\bibitem{TuraevViro:1992}
V. Turaev and O. Viro, {\em Topology} {\bf 31} (1992) 865.

\bibitem{Ooguri:1992}
H. Ooguri, {\em Nucl. Phys.} {\bf B382} (1992) 276; {\em Mod. Phys. 
Lett. A} {\bf 7} (1992) 2799.

\bibitem{Carlip:1998}
S. J. Carlip, {\em Quantum Gravity in 2+1 Dimensions},
Cambridge University Press (Cambridge, 1998). 

\bibitem{Barrett:1998}
J. W. Barrett and L. Crane, {\em J. Math. Phys.}
{\bf 39} (1998) 3296.

\bibitem{Roberts:1999}
%Justin Roberts, {\em Geometry \& Topology} {\bf 3} (1999) 21.
J. Roberts, {\em Geometry \& Topology} {\bf 3} (1999) 21.

\bibitem{FreidelLouapre:2003}
%Laurent Freidel and David Louapre, {\em Class. Quantum Grav.} 
L. Freidel and D. Louapre, {\em Class. Quantum Grav.} 
{\bf 20} (2003) 1267.

\bibitem{SchultenGordon:1975}
K. Schulten and R. G. Gordon, {\em J. Math. Phys.} {\bf 16} (1975) 
1961; {\bf 16} (1975) 1971.  

\bibitem{TaylorWoodward:2005}
%Yuka U. Taylor and Christopher T. Woodward, {\em Sel. Math., 
Y. U. Taylor and C. T. Woodward, {\em Sel. Math., 
NNew ser.} {\bf 11} (2005) 539.

\bibitem{Gurau:2008}
%Razvan Gurau, {\em Ann. H. Poincar\'e} {\bf 9} (2008) 1413.
R. Gurau, {\em Ann. H. Poincar\'e} {\bf 9} (2008) 1413.

\bibitem{HaggardLittlejohn:2010}
H.~M.~Haggard and R.~G.~Littlejohn
{\em Class. Quant. Grav.} {\bf 27} (2010) 135010.

\bibitem{Arnold:1989} 
V.~I.~Arnold V I 
{\it Mathematical Methods of Classical Mechanics} Springer-Verlag 
(New York, 1989).

\bibitem{Littlejohn:1990}
R.~G.~Littlejohn
{\em J. Math. Phys.} {\bf 31} (1990) 2952.
 
\bibitem{Aquilantietal:2007}
V.~Aquilanti, H.~M. Haggard, R.~G. Littlejohn, and L.~Yu, 
 {\em J. Phys. A} {\bf 40} (2007)  5637.

\bibitem{Wigner:1959}
E.~P.~Wigner, {\em Group Theory},
Academic Press (New York, 1959).

\bibitem{Kapovich:1996}
M. Kapovich and J. Millson, {\em J. Diff. Geom.} {\bf 44} (1996)  3.

\bibitem{Aquilantietal:2012}
V.~{Aquilanti}, H.~M. {Haggard}, A.~{Hedeman}, N.~{Jeevanjee}, R.~G.
  {Littlejohn}, and L.~{Yu}, 
   {\em J. Phys. A:Math. Theor.} {\bf 45} (2012) 065209.
   
\bibitem{LittlejohnYu:2009}
R.~G.~Littlejohn and L. Yu, {\em J. Phys. Chem. A} {\bf 113} (2009)
14904.

\bibitem{Littlejohn:1986}
R.~G.~Littlejohn, {\em Phys. Rep.} {\bf 138} (1986) 193.

\bibitem{Goldstein:1980}
H.~Goldstein, {\em Classical Mechanics} 2nd ed, Addison Wesley
(Reading, Mass., 1980).
   
\bibitem{PletyukhovEtal:2002}
M. Pletyukhov, Ch. Amann, M. Mehta and M. Brack, {\em Phys. Rev. Lett.} 
{\bf 89} (2002) 116601.

\bibitem{GargStone:2004}
%Anupam Garg and Michael Stone, {\em Phys. Rev. Lett.} {\bf 92} 
A. Garg and M. Stone, {\em Phys. Rev. Lett.} {\bf 92} 
(2004) 010401.

\end{thebibliography}
\end{document}